\documentstyle[psfig]{mn}

\newif\ifAMStwofonts

\newcommand{\msol}{\hbox{\,${\rm M_{\sun}}$}}

\newcommand{\cmsq}{\hbox{\,${\rm cm}\,{\rm s}^{-2}$}}

\newcommand{\cmc}{\hbox{\,${\rm cm^{-3}}$}}
\newcommand{\kms}{\hbox{\,${\rm km\,s^{-1}}$}}

\newcommand{\SS}{{Section}\ }
\newcommand{\etal}{{et~al.}}
\newcommand{\eg}{{e.g.}\ }
\newcommand{\cf}{{cf.}\ }
\newcommand{\ie}{{i.e.}\ }

\newcommand{\dpi}{\hbox{$\delta \Pi$}}
\newcommand{\ap}{\hbox{$a^{\prime}$}}
\newcommand{\psiz}{\hbox{$\Psi_{0}$}}
\newcommand{\grad}{\hbox{$g_{rad}$}}
\newcommand{\gz}{\hbox{$g_{rad}^0$}}
\newcommand{\gpul}{\hbox{$g_{pull}$}}

\newcommand{\rmi}{\hbox{r$_{\rm min}$}}

\newcommand{\uz}{\hbox{$\rm{U}_0$}\/}

\newcommand{\nbar}{\hbox{$\bar n$}}

\newcommand{\qz}{\hbox{q$_0$}}
\newcommand{\nz}{\hbox{$n_0$}}

\newcommand{\mud}{\hbox{$\mu_{\rm D}$}}

\newcommand{\av}{\hbox{${\rm A}_{\rm V}$}}
\newcommand{\ev}{\,{\sl eV}\/}
\newcommand{\kev}{\,{\sl keV}\/}
\newcommand{\mi}{\,$\mu$m\/}

\newcommand{\fexiw}{\hbox{[Fe\,{\sc xi}]\,$\lambda $7892}}
\newcommand{\fex}{\hbox{[Fe\,{\sc x}]}}
\newcommand{\fexw}{\hbox{[Fe\,{\sc x}]\,$\lambda $6374}}
\newcommand{\fevii}{\hbox{[Fe\,{\sc vii}]}}
\newcommand{\feviiw}{\hbox{[Fe\,{\sc vii}]\,$\lambda $6086}}

\newcommand{\neviw}{\hbox{[Ne\,{\sc vi}]\,7.66$\mu$m}}
\newcommand{\nevbw}{\hbox{[Ne\,{\sc v}]\,14.32$\mu$m}}
\newcommand{\nevcw}{\hbox{[Ne\,{\sc v}]\,24.31$\mu$m}}

\newcommand{\nev}{\hbox{[Ne\,{\sc v}]}}

\newcommand{\mgvw}{\hbox{[Mg\,{\sc v}]\,5.62$\mu$m}}

\newcommand{\mgviiw}{\hbox{[Mg\,{\sc vii}]\,5.51$\mu$m}}

\newcommand{\mgviiiw}{\hbox{[Mg\,{\sc viii}]\,3.03$\mu$m}}

\newcommand{\oivw}{\hbox{[O\,{\sc iv}]\,25.90$\mu$m}}

\newcommand{\oii}{\hbox{[O\,{\sc ii}]}}

\newcommand{\oi}{\hbox{[O\,{\sc i}]}}

\newcommand{\sivi}{\hbox{[Si\,{\sc vi}]}}
\newcommand{\siviw}{\hbox{[Si\,{\sc vi}]\,1.963$\mu$m}}

\newcommand{\siviiw}{\hbox{[Si\,{\sc vii}]\,2.483$\mu$m}}
\newcommand{\siix}{\hbox{[Si\,{\sc ix}]}}
\newcommand{\siixw}{\hbox{[Si\,{\sc ix}]\,3.935$\mu$m}}

\newcommand{\sii}{\hbox{[S\,{\sc ii}]}}

\newcommand{\sivw}{\hbox{[S\,{\sc iv}]\,10.54$\mu$m}}

\newcommand{\sviiiw}{\hbox{[S\,{\sc viii}]\,$\lambda $9913}}

\newcommand{\sixw}{\hbox{[S\,{\sc ix}]\,1.25$\mu$m}}
\newcommand{\caviii}{\hbox{[Ca\,{\sc viii}]}}
\newcommand{\caviiiw}{\hbox{[Ca\,{\sc viii}]\,2.32$\mu$m}}

\ifoldfss
  \ifCUPmtlplainloaded \else
    \NewTextAlphabet{textbfit} {cmbxti10} {}
    \NewTextAlphabet{textbfss} {cmssbx10} {}
    \NewMathAlphabet{mathbfit} {cmbxti10} {} 
    \NewMathAlphabet{mathbfss} {cmssbx10} {} 
  \fi
  \ifAMStwofonts
    \ifCUPmtlplainloaded \else
      \NewSymbolFont{upmath} {eurm10}
      \NewSymbolFont{AMSa} {msam10}
      \NewMathSymbol{\upi}     {0}{upmath}{19}
      \NewMathSymbol{\umu}     {0}{upmath}{16}
      \NewMathSymbol{\upartial}{0}{upmath}{40}
      \NewMathSymbol{\leqslant}{3}{AMSa}{36}
      \NewMathSymbol{\geqslant}{3}{AMSa}{3E}

       \let\le=\leqslant
       \let\ge=\geqslant
    \fi
  \fi
\fi 

\ifnfssone
  \newmathalphabet{\mathit}
  \addtoversion{normal}{\mathit}{cmr}{m}{it}
  \addtoversion{bold}{\mathit}{cmr}{bx}{it}
  \newmathalphabet{\mathbfit} 
  \addtoversion{normal}{\mathbfit}{cmr}{bx}{it}
  \addtoversion{bold}{\mathbfit}{cmr}{bx}{it}
  \newmathalphabet{\mathbfss} 
  \addtoversion{normal}{\mathbfss}{cmss}{bx}{n}
  \addtoversion{bold}{\mathbfss}{cmss}{bx}{n}
  \ifAMStwofonts
    \ifCUPmtlplainloaded \else
      \UseAMStwoboldmath
      \makeatletter
      \new@mathgroup\upmath@group
      \define@mathgroup\mv@normal\upmath@group{eur}{m}{n}
      \define@mathgroup\mv@bold\upmath@group{eur}{b}{n}
      \edef\UPM{\hexnumber\upmath@group}
      \new@mathgroup\amsa@group
      \define@mathgroup\mv@normal\amsa@group{msa}{m}{n}
      \define@mathgroup\mv@bold\amsa@group{msa}{m}{n}
      \edef\AMSa{\hexnumber\amsa@group}
      \makeatother
      \mathchardef\upi="0\UPM19
      \mathchardef\umu="0\UPM16
      \mathchardef\upartial="0\UPM40
      \mathchardef\leqslant="3\AMSa36
      \mathchardef\geqslant="3\AMSa3E

       \let\le=\leqslant
       \let\ge=\geqslant
    \fi
  \fi
\fi 

\ifnfsstwo
  \DeclareMathAlphabet{\mathbfit}{OT1}{cmr}{bx}{it}
  \SetMathAlphabet\mathbfit{bold}{OT1}{cmr}{bx}{it}
  \DeclareMathAlphabet{\mathbfss}{OT1}{cmss}{bx}{n}
  \SetMathAlphabet\mathbfss{bold}{OT1}{cmss}{bx}{n}
  \ifAMStwofonts
    \ifCUPmtlplainloaded \else
      \DeclareSymbolFont{UPM}{U}{eur}{m}{n}
      \SetSymbolFont{UPM}{bold}{U}{eur}{b}{n}
      \DeclareSymbolFont{AMSa}{U}{msa}{m}{n}
      \DeclareMathSymbol{\upi}{0}{UPM}{"19}
      \DeclareMathSymbol{\umu}{0}{UPM}{"16}
      \DeclareMathSymbol{\upartial}{0}{UPM}{"40}
      \DeclareMathSymbol{\leqslant}{3}{AMSa}{"36}
      \DeclareMathSymbol{\geqslant}{3}{AMSa}{"3E}

       \let\le=\leqslant
       \let\ge=\geqslant
    \fi
  \fi
\fi 

\ifCUPmtlplainloaded \else
  \ifAMStwofonts \else 
    \def\upi{\pi}
    \def\umu{\mu}
    \def\upartial{\partial}
  \fi
\fi

\title[Coronal Gas in Active Galaxies]{Radiative Acceleration of
Coronal Gas in Seyfert Nuclei}

\author[L. Binette]{L. Binette \\
European Southern Observatory, Casilla 19001, Santiago 19, Chile
(E-mail: lbinette@eso.org) }

\date{Accepted 18 December 1997.~ Received 27 August 1997 }

\pagerange{\pageref{firstpage}--\pageref{lastpage}}
\pubyear{1994}

\begin{document}

\maketitle

\label{firstpage}

\begin{abstract}

The line ratios from coronal gas in Seyferts can be successfully
fitted with photoionized clouds of high densities and low volume
filling factor. The ionization parameter implied is sufficiently high
that models must consider the effect of radiation pressure from the
active nucleus.  In spite of the nucleus gravitational force,
radiation pressure is sufficiently strong to compress and radially
accelerate the internally stratified gas clouds provided these contain
small amounts of dust ($\simeq 10$\% of solar neighborhood
value). This radial acceleration could explain the blueshift of the
coronal lines relative to systemic velocity without the need of
invoking an ambient `pushing' wind. Embedded dust has the
interesting effect of making the photoionized clouds marginally
ionization bounded instead of matter bounded.
\end{abstract}


\section{Introduction} \label{intro}

Studies of the emission lines in Seyfert galaxies have indicated the
presence of two different regions: a region of very high density with
large velocity spread -- the broad line region (BLR), and a region of
low gas density with a much smaller velocity spread -- the narrow line
region (NLR). However, there is also a set of forbidden lines of
extremely high excitation, the so-called `coronal' lines which
includes for instance \feviiw, \fexw, \sviiiw, \siviw\ and
\siviiw. Most models favor photoionization over collisional ionization
as excitation mechanism (\eg\ Korista \& Ferland 1989; Oliva \etal\
1994; Moorwood \etal\ 1996). Earlier models considered that the
filling factor of the coronal gas approached unity with a density
lower than that of the NLR. The advent of far infrared measurements of
the density sensitive \nev\ 14.3\mi/24.3\mi\ line ratio with ISO
(Moorwood \etal\ 1996: Mo96), however, suggests much higher densities
of $\sim 5\,000$\cmc\ implying a very low volume filling factor for
the coronal gas. In Paper~I, Binette \etal\ (1997) proposed that these
lines originate from individual gas clouds -- as for the NLR -- but
characterized by an unusually high ionization parameter, \uz\ [\cf\
equation~(1) below]. A similar gas geometry underlies the extensive
grid of photoionization models of Ferguson, Korista \& Ferland 
(1997). In Paper~I, the clouds responsible for the coronal lines are
always matter-bounded while they are essentially ionization-bounded in
the models of Ferguson \etal\ (1997) or in the photoevaporing dusty
cloud model of Pier \& Voit (1995). One advantage of the
matter-bounded clouds is that they are easier to accelerate by
radiation pressure owing to their lower mass. Since the coronal lines
are known to be systematically blueshifted in Seyfert galaxies
(Penston \etal\ 1984) relative to the systemic velocity (by 35\kms\ in
the case of Circinus, see Oliva \etal\ 1994; hereafter OSMM), our
justification of using matter-bouneded clouds was to account for this
property by incorporating self-consistently the effect of radiation
pressure in our cloud model.

The ionization parameters required to fit the very high excitation
coronal lines is very high, \uz$\ga 0.2$, implying that the pressure
exerted by the ionizing radiation exceeds the gas pressure (see
Paper~I). This has two interesting consequences: first, the
photoionized cloud can be radially accelerated until the ram presure
it exerts on the ambient medium equals the pressure of the latter,
this would account for the blueshift of the coronal lines for certain
geometries, second, a density gradient arises within the clouds due to
radiation pressure (a distributed force) and as shown in Paper~I, the
density stratification generated in this way leads to a wider range in
excitation of the ionization species present in the cloud. The
question addressed in this Letter is whether the pressure from the
nuclear radiation field is sufficient to cause a strong radial
acceleration of the clouds despite the gravitational pull exerted by
the stars and the nuclear black hole.

\section{The model} \label{model}


The multipurpose code {\sc mappings~ic} was used to compute the
photoionization models. The atomic data is taken from a compilation of
Ralph Sutherland (\cf\ Appendix in Ferruit \etal\ 1997). The
uncertainties in collision strengths of many coronal lines are often
large (see review by Oliva 1997 and the discussion of the `iron
conundrum' by FKF) and may dominate the errors in the predicted line
strengths.

\subsection{Gas abundances and dust content \mud}  \label{depl}

As is customary and in the absence of other information, the
abundances adopted will be solar (Anders \& Grevesse 1989) but
depleted according to the dust content. The latter is defined by the
quantity \mud\ which is the dust-to-gas ratio of the plasma expressed
in units of the solar neighborhood dust-to-gas ratio. The effects of
dust on the thermal and ionization structure are taken into account
(Binette \etal\ 1993). However, a different depletion scheme is
defined whereby the destruction of dust grains is assumed to return
uniformly the depleted elements (`uniform return'). For instance,
supposing \mud=0.1 (i.e. 10\% of the solar neighborhood dust content),
the gas phase abundances are derived from the fully depleted set of
abundances (\mud$\equiv$1) to which we restitute 90\% of what has been
depleted. This means that even for the heavily depleted elements like
Ca, the gas phase abundance when \mud=0.1 is 90\% solar. The
justification behind this is that small values of \mud\ are not the
result of a different dust formation history but rather that of recent
destruction (of normally formed dust) by sputtering, sublimation or
optical erosion.

\subsection{The ionizing continuum} \label{uvco}

We adopt a similar ionizing continuum as in Paper~I, namely a powerlaw
of index $\gamma =-1.3$ ($\varphi_{\nu} \propto \nu^{+\gamma}$) for
the infrared-UV domain. This powerlaw joins at 2000\ev\ with a
flatter powerlaw of index $\gamma =-0.7$ to cover the X-ray
domain. We impose a high energy cut-off of 100\,\kev (see Mathews \&
Ferland 1987).
As is customary, we define the ionization parameter \uz\ as the ratio
between the density of impinging ionizing photons and the gas density
at the face of the cloud
\begin{equation}
{\rm U_0} = \frac{\int_{\nu_1}^{\infty}{\varphi_{\nu}d\nu/h\nu}}{c \,
n_0} = \frac{{\rm q}_{0}}{c \, n_0} \;,
\end{equation} 
where $\varphi_{\nu}$ is the monochromatic energy flux impinging on
the cloud, $\nu_1$ the Lyman frequency, \qz\ the number of ionizing
photons incident on the slab per cm${^2}$ per second, $c$ the speed of
light and $n_0$ the total gas density at the irradiated surface of the
slab (all quantities relevant to the irradiated surface carry $0$ as a
superscript or subscript).

\subsection{Acceleration and pressure stratification} \label{acce}

We adopt the formalism developed by Mathews (1986) and consider that
our radially accelerated cloud has achieved internal hydrostatic
equilibrium in the radial (outward) direction in the noninertial
accelerating frame. Therefore, across a cloud (hereafter approximated as
a slab) of total column density $N_t$ and situated at a distance $r$
from the black hole (BH), we have
\begin{equation}
{{1}\over{m_H}}{{dP}\over {dN}}= g_{rad}(N) - (g_{pull} + a_{cl}) = g_{rad}(N) -a^{\prime} \;, \label{eqdi}
\end{equation}
where $m_{H}$ is the proton mass, $g_{rad}(N)$ the local {\it
radiative} acceleration due to radiation pressure, \gpul\ the
gravitational acceleration due to the central massive BH and the
inner stars, and $a_{cl}$ the instantaneous acceleration of the entire
cloud. The volume force exerted by radiation pressure is simply
$F_{rad} = n m_{H} g_{rad}(N)$ with its numerical expression in
terms of photoelectric cross sections defined in Appendix A.3 of
Paper~I.  The clouds geometrical thickness is considered negligible as
compared to nuclear distance $r$, a good approximation in the current
context of moderately thick clouds. This implies that for a cloud at
$r$, the modified acceleration $a^{\prime}\; ~(= g_{pull} + a_{cl})$
is constant across its thickness $N_t$.  We can regard $a^{\prime}$ as
an eigenvalue for the acceleration of the cloud.
In our scheme, the systematic shift\footnote{It is believed that we
see one hemisphere only (\ie\ one ionization cone) of the NLR 
in both Circinus and NGC~1068.} towards the blue of the lines requires
that $a^{\prime} > g_{pull}$, that is $a_{cl}>0$ in equation~(2). The
blueshift can either be the result of an earlier acceleration of the
clouds (the clouds have reached their terminal velocity against the
opposing drag and gravitational forces), or of ongoing cloud
acceleration.

\subsection{Implementation in {\sc mappings~ic}} \label{impl}

Independently of whether the acceleration corresponds to bulk motion
or to inertial acceleration, $a^{\prime}$ is an important parameter
for the cloud's density structure. In our model, we express this input
parameter in terms of the dimensionless quantity $\Psi_0$ defined as
$\Psi_0 = a^{\prime}/g_{rad}^0$ with \gz\ the radiative acceleration
at the irradiated face.

As described in Appendix of Paper~I, MAPPINGS integrate equation~(2) as
well as simultaneously solve for the local ionization and thermal
balance as a function of depth $N$.  If \uz\ is sufficiently high,
equation~(2) implies a positive pressure gradient within the slab
therefore leading to a strong density gradient across the photoionized
slab. We define the relative pressure difference between the
irradiated face and the back of the slab, $\delta \Pi$, as
\begin{equation}
\delta \Pi = \left(P_{gas}^{back}-P_{gas}^0 \right) / {\rm MIN}
\left( P_{gas}^{back},P_{gas}^0 \right)  \label{pie} \;,
\end{equation}
where the denominator represents the isotropic pressure, $\widehat P$,
of the ambient cloud-confining medium. Note that \dpi\ is approximately
the ratio of dynamical pressure to ambient pressure. If $\delta \Pi
<0$, the cloud is experiencing a dynamical pressure at the inner
irradiated surface (`pushed' clouds by an ambient wind), while if
$\delta \Pi >0$ the clouds are suffering a drag force at their outer
back surface as they are `pushing' against the ambient medium. All the
models presented below correspond to this latter case of $\delta \Pi
>0$.

For the dust-free matter bounded slab discussed in Paper~I where we
neglected inertial and bulk acceleration ($\Psi_0\equiv 0$), radiative
pressure was shown to induce a pressure difference in the slab as high
as $\delta \Pi =1.6$ when \uz=0.5\,. Since any realistic nuclear
environment will imply a nonzero $a^{\prime}$ as a result of the
gravitational field, it is the object of this study to investigate the
effect of a positive \psiz.

In the general case of $\Psi_0 >0$, because $g_{rad}$ decreases
monotonically with depth as the impinging radiation is gradually
absorbed, the right term of equation~(2) become negative
at very large depths.  This can result in $\delta \Pi <0$ depending on
the choice of $N_t$, \mud\ and $\Psi_0$. Although we could interpret
this as being caused by ram pressure of an outflowing ambient medium
(`pushed' cloud), this domain of parameters has not been explored:
first, because the aim of the Letter is to relate the line blueshift
to the acceleration caused by radiation pressure and not to an
extraneous alternative mechanism, second, because all the models
considered below are not sufficiently thick to obtain \dpi$<$0.

To maximize the acceleration \ap, it is useful to increase \psiz\ as
much as possible. However, it was found that \psiz=0.5 which is
adopted hereafter cannot be exceeded significantly (otherwise $\delta
\Pi$ shrinks considerably). In effect, \psiz=0.5 maintains two
essential effects: it still allows a significant density gradient
(therefore optimizing the line ratio fit, \cf\ Paper~I) and it
results at the same time in a strong acceleration $a^{\prime}$.
Clouds submitted to significant radiation pressure are particularly
prone to derimming through the establishment of a lateral flow as
shown by Mathews (1986). For a review of the problem of cloud
stability under conditions appropriate to the NLR, see Mathews \&
Veilleux (1989).

\section{The excitation of the coronal gas in Circinus and NGC\,1068} \label{exc}

All detailed studies of the coronal gas reveal that the emitting gas
encompasses many stages of excitation. This indicates that this gas
must be significantly thick to the ionizing radiation rather than very
optically thin. We can envisage two geometries, one broadly spherical
and filled with low density gas: the onion ring geometry (Korista \&
Ferland 1989; OSMM) the other consists of an ensemble of clouds of
very low filling factor and therefore similar to the NLR gas in
general except for its much higher excitation. What is clear is that
although the range of ionization species of the coronal gas is wide,
it does not encompass the very low excitation species of \oii, \oi,
\sii... since the latter appear to be kinematically distinct as shown
by OSMM and Marconi \etal\ (1996). In Paper~I, this was interpreted as
an indication that the coronal line emitting clouds were matter
bounded as was also thought to be the case for the high excitation
clouds of the ENLR and NLR \cite{bwsb,wbsb}.

The thickness of the clouds is not defined arbitrarily but can be
empirically set by the ratio between the highest and the lowest
excitation species belonging to the coronal gas.  In Paper~I it was
found that the lines of \siviw\ and \siixw\ worked very well for this
purpose. The same procedure is adopted here whereby the slab is
truncated at the point where the modeled \sivi/\siix\ ratio matches
the observed value. It is interesting to note that in the {\it
dust-free} case, despite the difference in excitation between the two
Seyferts, this truncation always seems to occur after $\sim 60$--65\%
of the ionizing photons have been {\it absorbed}.

\begin{figure} 
\centerline{\psfig{figure=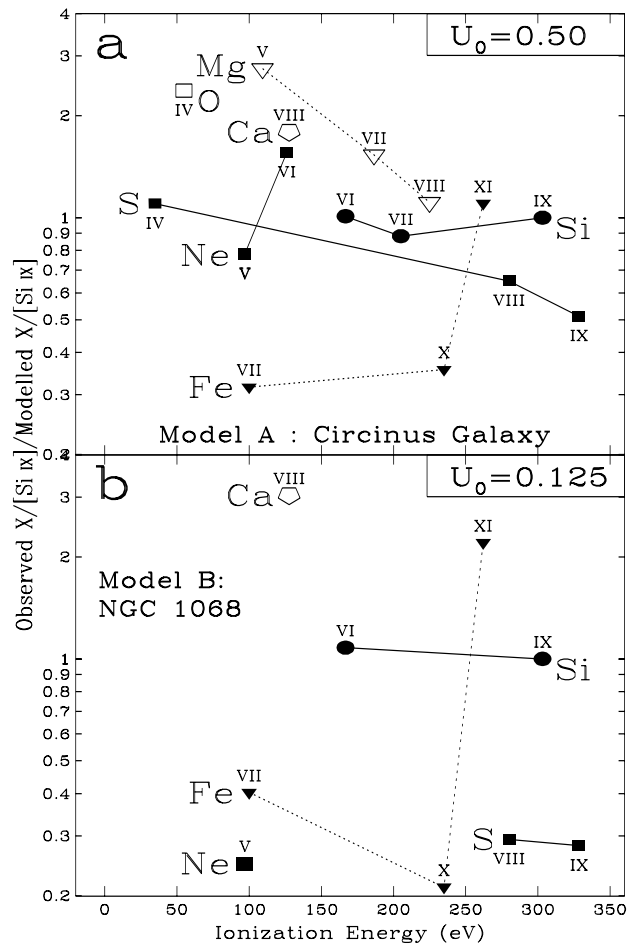,height=14cm,width=8.7cm}} 
\caption{Observed divided by modeled line flux ratios for (a) the
Circinus galaxy and (b) NGC\,1068. All ratios are expressed relative
to \siixw. Both models A and B have \nz=1200\,\cmc\ and a dust content
of 10\% of the solar neighborhood dust-to-gas ratio (\mud=0.1, see
text and Table.~1). The lines plotted in (a) include ~\oivw, ~\siviw,
~\siviiw, ~\siixw, ~\sivw, ~\sviiiw, ~\caviiiw, ~\sixw,~\nevbw,
~\nevcw, ~\neviw, ~\feviiw, \fexw, \fexiw, ~\mgvw, \mgviiw\ and
~\mgviiiw.
\label{figcif} }
\end{figure}

\begin{enumerate}

\item{The Circinus galaxy.} \label{circ} 

If we repeat the dust-free calculations of Paper~I with \uz=0.5 and
density \nz=1200\,\cmc\ but set \psiz=0.5, we find that the modified
acceleration is only \ap$=2.9 \times 10^{-7} $\,\cmsq\ (\ie\ \gz\ is
twice that value). As shown later in \SS~\ref{mass}, this is at least
an order of magnitude below the acceleration due to the gravity
exerted by the stars within 100\,pc of the nucleus of the
Milky~Way. It would therefore appear that radiation pressure is
insufficient in accelerating radially the line emitting gas. However,
if we allow trace quantities of dust to be present in the cloud,
radiation pressure is much higher and dust absorption even dominates
the photoelectric absorption terms contained in \grad. For instance,
if we set \mud=0.1, we obtain \ap$=2.6 \times 10^{-6}\,$\cmsq. The
modeling of the observed line ratios reported by Mo96 and OSMM is
presented in Fig.~1a (Model~A) using \mud=0.1\,. The fit is of
comparable quality to that presented in Paper~I or in Mo96. As can be
expected, the effect of `uniform return' depletion on the \caviii\
point is less than 0.05\,dex (see \SS~\ref{depl}). The resultant
pressure stratification is somewhat smaller at \dpi=1.2\,.  It was
found that arbitrarily increasing either \mud\ or \psiz\ to higher
values degraded significantly the fit to the Circinus line ratios.

The hypothesis of internal dust is also present in the comprehensive
dusty cloud model of Pier \& Voit (1995). In their model, the strong
radiation pressure exerted on the dusty gas (situated near the inner
molecular torus) leads to progressive cloud evaporation and to the
creation of an X-ray-heated wind. The smaller dust content and cloud's
thickness of the matter-bounded clouds presented in this Letter leads
to a smaller extinction (\av$\simeq 0.15$) than in Pier \& Voit
(1995).

\item{The Seyfert galaxy NGC\,1068.} \label{ngc} 

We follow a similar procedure in the case of NGC\,1068. The
observational data is taken from Marconi \etal\ (1996). The set of
lines is not as complete but is probably sufficient for estimating
\uz. Marconi \etal\ reported that the excitation was lower than in
Circinus.  A similar result is found here. A value four times lower
than in Circinus leads to a reasonable fit to the gas {\it excitation}
as revealed by the S and Si line ratios. The fit is shown in
Fig.~1b (Model~B: \uz=0.125 and \mud=0.1). Systematic but equal
departures from unity (\eg\ the S lines) might partly be due to
peculiar abundances relative to solar. In this model, we arbitrarily
adopted the same density as for Circinus since no published
measurements of the density from the \nev\ 14.3\mi/24.3\mi\ ratio was
available at the time of writing. The excitation indicated by the
ratio \fevii/\fex\ is not so badly fitted but \fexiw\ is certainly
discrepant. As was the case for Circinus, assigning more than 50\% of
the radiation pressure to cloud acceleration (\ie\ \psiz$>$0.5) was
found to worsen the fit. On the other hand, the amount of dust allowed
did not appear to strongly affect the fit as long as \mud$\le 0.5$. A
good fit to the value of \ap\ as a function of the dust content is~
$a^{\prime} = ({n_0}/1200) ({\rm U}_0/0.125) \left[3.1 \times 10^{-7}
+ 5.8 \times 10^{-6} \mu_{D} \right]$.

\end{enumerate}

\begin{table*}
\centering
\begin{minipage}{180mm}
\caption{Parameters characterizing the models.}
\label{tbl-1}
\begin{tabular}{@{}cccccccccccccc}
{Model} & {\uz }   & 
{\nz }   & {\mud } & 
{$N_t/10^{21}$}  & {\nbar } & 
{${\rm A}_{\rm V}$} & 
{\dpi }     & {\ap/$10^{-6}$ }  & 
{\nbar/\nz }   & 
{${\rm q}_{back}/{\rm q}_{0}$\footnote{Fraction of ionizing photons escaping the slab unabsorbed.}}  &
{$g_{dyn}/g_{rad}$\footnote{Ratio of dynamical acceleration to total radiative acceleration (Mathews 1986).}} &
{${\widehat P}/k/10^7K$\footnote{Isotropic pressure of the ambient
cloud-confining medium. It would scale with \nz\ if \uz\ is kept constant.}} &
{L~(pc)\footnote{Geometrical depth of the slab in parsecs. L would scale
inversely with \nz\ if \uz\ is kept constant. }}   \\
 
A & 0.50  & 1200 & 0.10 & 6.2 & 4500 & 0.15 & 1.22 & 2.63 & 3.75 &
0.06 & $-$0.36 & 13.2 & 0.68 \\
 
B & 0.125 & 1200 & 0.10 & 3.0 & 2100 & 0.07 & 0.51 & 0.89 & 1.75 &
0.22 & $-$0.46 & 7.8 & 0.55 \\
 
C & 0.125 & 1200 & 0.30 & 2.6 & 2000 & 0.19 & 0.25 & 2.05 & 1.65 &
0.04 & $-$0.18 & 8.1 & 0.35  \\
\end{tabular}
\end{minipage}
\end{table*}

If we vary \nz\ and keep the ionization parameter constant,
by equation~(1) \qz\ must scale with \nz. Therefore the modified
accelerations \ap\ will also scale linearly with \nz\ since radiation
pressure scales with ionizing flux. Note that a similar model to that
of Fig.~1b but for which the dust content has been increased
to \mud=0.3 (i.e. Model~C in Table~1), the acceleration \ap\ becomes
comparable to that of Model~A with \uz\ four times larger
(but three times less dust).

For Models~A and B, the contrast between the density at the irradiated face
and the `emissivity averaged' value, \nbar/\nz, is 3.75 and 1.75
respectively. Table~1 list other relevant quantities for the two
models shown in Fig.~1 as well as for a dustier Model~C characterized
by \mud=0.3.

\section{Results} \label{resu}

For an empirically determined value of \uz, \ap\ will scale linearly
with the ionizing flux \qz. Therefore, for a given value of \uz, the
predicted value of \ap\ scales linearly with the assumed value of the
density \nz. Using the \nev\ 14.3\mi/24.3\mi\ line ratio, one can
determine the density of the coronal line region and therefore \ap. It
is found that the mean density \nz\ as defined in Paper~I for a
pressure stratified slab is the same as that inferred from the
calculated far-infrared \nev\ line ratio. Therefore this line ratio
reveals us the `characteristic' density of the highly excited
component. Having determined empirically the ionization parameter
characterizing the coronal gas in two AGN, we can now proceed to
compare the calculated \ap\ of models with, \gpul, the strength of
the gravitational field expected in a galactic nucleus.

\subsection{The nuclear mass model: $M_{\star+bh}(r)$ } \label{mass}

\begin{figure}
\centerline{\psfig{figure=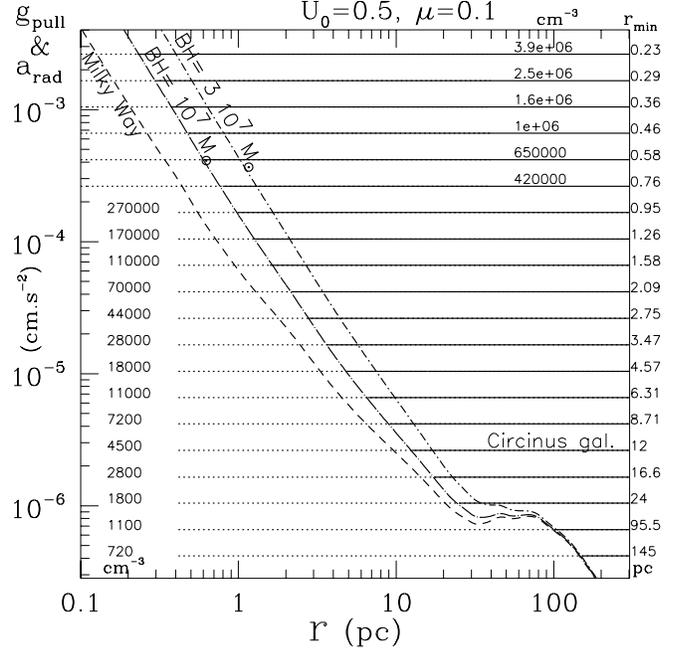,height=8.7cm,width=8.7cm}}
\caption{A plot of the gravitational pull, $g_{pull}$, exerted by the
nuclear mass distribution of the Milky Way (dash line) as a function
of galactic radius. The continuous and the {\it dash-dotted} lines
consider the effect of a more massive BH of $10^7$ or $3\times
10^7$\msol, respectively (see \SS~\ref{mass}).  The modified
acceleration $a^{\prime}$ due to radiation pressure for clouds of
different densities \nbar\ is indicated using horizontal lines
(whatever \nbar, it is assumed that \uz=0.5).  The minimum radius for
which coronal gas emission can be accelerated (i.e. $a^{\prime} \ge
g_{pull}$) is indicated in the right margin for the case of a
$10^7$\msol\ BH. \label{facir} }
\end{figure}

For radiative acceleration to be responsible of the blueshift of the
coronal lines, it must overcome the gravitational pull of the combined
mass of the central stars and the nuclear BH. To illustrate the force
entailed, let us adopt the better studied Milky Way mass distribution,
$M_{\star+bh}(r)$, inside a radius $r \le 200$\,pc. From 0.1 to 2\,pc
we borrow the mass model of Genzel \etal\ (1997) which we join at
2\,pc to the isotropic orbit stellar mass model of Lindqvist \etal\
(1992; \cf\ their Table~3) which is based on the projected mass method
for a dominant central mass. The resultant gravitational pull is
simply $g_{pull}=G M_{\star+bh}/r^2$ and is plotted in
Fig.~\ref{facir} (dash line). One interpretation proposed by Genzel
\etal\ for the innermost mass is that it might consist of a BH of
$\approx 2.8 \times 10^{6} M_{\sun}$. Note that beyond $r \ga 30$\,pc
the gravitational pull from the stars dominate that of the BH. On
account of the much stronger nuclear activity characterizing Seyferts,
it is plausible that a typical AGN could have a more massive BH than
the Milky\,Way. For instance, estimates of the central mass in
NGC\,1068 ranges from $10^{7} M_{\sun}$ \cite{grgw} to $3 \times
10^{7} M_{\sun}$ \cite{gaet}. In an attempt to allow for this
possibility, the effect of simply substituting a more massive point
source is shown in Fig.~\ref{facir} in which the continuous and the
dotted-dash curves correspond to BH masses of $10^{7} M_{\sun}$ and $4
\times 10^{7} M_{\sun}$, respectively.

\subsection{Minimum NLR radius} \label{mini}

It is not possible at this point to realistically assess how different
the nuclear stellar field would be in either more massive galaxies
than ours or in nuclei with more massive BHs. With this caveat in
mind, let us adopt the continuous curve in Fig.~\ref{facir} of a
$10^{7} M_{\sun}$ BH as being representative of low luminosity
AGN. After determining the different values of \ap\ expected if we
simply vary \nz\ of Model~A (with \uz=0.5), we can proceed to derive
the minimal radius beyond which the clouds have their acceleration
\ap\ exceed $g_{pull}$. The allowed domain in $r$ for uniform
acceleration of the coronal gas lies to the left of the $g_{pull}$
curve in Fig.~\ref{facir} as shown by the horizontal {\it solid}
lines, each corresponding to different values of the mean densities
\nbar.  For the particular case of the Circinus galaxy for which we
know the density, a minimum radius of 12\,pc\ is inferred inside of
which the coronal gas would fall or stall instead of being radially
accelerated. The true position of the nucleus in Circinus is not known
with precision since its nucleus is heavily reddened and the true
kinematic centre may not correspond to the continuum peak. We cannot
yet therefore verify the validity of \rmi. However, in the case of
NGC\,1068, Marconi \etal\ (1996) found that the infrared lines were
displaced by 0.5\arcsec\ from the optical continuum peak, which
corresponds to a nuclear offset of $\simeq 40$\,pc (D=18\,Mpc). The
gas density is not yet determined but if comparable to Circinus, we
would expect a much larger \rmi\ in NGC\,1068 based on its lower
excitation by a factor four.

The main conclusions are the following. Provided there is dust left
out in the gas, radiation pressure not only can induce a density
gradient in the photoionized clouds but also be responsible for their
outward acceleration despite the presence of strong gravitational
field.  It is found that higher values than \mud=0.1 are detrimental
to the fit in the case of Circinus.  Based on empirically measured
densities and excitation, a minimum nuclear distance \rmi\ is
predicted to characterize the {\it observable} coronal gas. The higher
the density, the smaller \rmi. The advent of high spatial resolution
line mapping will allow testing whether such a correspondence between
$r$ and \nbar\ exists.
Finally, because of the large value of \uz, the fraction of ionizing
photons escaping the clouds can be quite small even with small amounts of
dust. By empirically setting the thickness of the clouds
to be such that the observed \sivi/\siix\ ratio is reproduced by the
model, one finds that only 6\% of the impinging photons in Model~A are
{\it not} absorbed (instead of 35--40\% in the dust-free case). One
may therefore conceive the possibility that the coronal clouds are not
matter bounded but marginally thick and bounded by dust
absorption. The most inner nuclear regions (presumably of higher
visibility in Seyferts~1) which still emit coronal lines may
correspond to the missing domain between the BLR and the NLR (this
gas would have to be proportionally denser to have $a^{\prime}
\ge g_{pull}$). In effect, as shown by Netzer \& Laor (1993), dust
embedded photoionized gas of high excitation presents a greatly
reduced line efficiency explaining the relative line weakness of this
putative intermediate region.


\bsp

\label{lastpage}

\end{document}